\documentclass{ws-procs10x7}
\usepackage{epsfig,graphicx}

\newcolumntype{d}[1]{D{.}{.}{#1}}

\newcommand{\bea}{\begin{small}\begin{eqnarray}}
\newcommand{\eea}{\end{eqnarray}\end{small}}
\newcommand{\upmns}{{U_{\text{PMNS}}}}
\newcommand{\udagpmns}{{U^\dagger_{\text{PMNS}}}}
\newcommand{\ustapmns}{{U^*_{\text{PMNS}}}}

\makeindex
\begin{document}

\title{MINIMAL LEPTON FLAVOUR VIOLATION AND LEPTOGENESIS}

\author{V. PORRETTI}

\address{Departament de F\'{\i}sica Te\`orica, Universitat de Val\`encia, E-46100, Burjassot, Spain and\\
Dipartimento di Fisica `E.~Amaldi', Universit\`a di Roma Tre and
INFN, Sezione di Roma Tre, I-00146 Rome, Italy}


\twocolumn[\maketitle\abstract{We study the role of leptogenesis in
the framework of minimal lepton flavour violation. }
]

\section{Outline}

If neutrino masses are generated by a see-saw mechanism with heavy
Majorana neutrinos, the hypothesis of minimal flavour violation
(MFV) in the lepton sector cannot be a simple replica of MFV in the
quark sector (sec.\ref{def}): more parameters are involved
(sec.\ref{count}) and high energy processes like leptogenesis
(sec.\ref{viable}) can have a role in the predictions for low energy
experiments (sec.\ref{effect}). We conclude summarizing the main
points in sec.\ref{sum}.

The introduction in sec.\ref{def} is mainly based on
Refs.\cite{D'Ambrosio:2002ex,Cirigliano:2005ck}, the rest on
Ref.\cite{Cirigliano:2006nu}. Aspects of MFV in the lepton sector
have been discussed also in
Refs.\cite{Cirigliano:2006su,Davidson:2006bd,Grinstein:2006cg,Branco:2006hz}.

\section{A definition of minimal lepton flavour violation}\label{def}

In the Standard Model (SM) flavour changing neutral currents are
absent at tree level and mixing between different families is
suppressed by small fermion masses and small CKM angles. Any new
physics (NP) that fails to reproduce these special features of the
SM is likely to give visible contributions to flavour changing
rates. Experiments indicate that this is not the case, so that or
flavour violating NP intervenes at a very high energy scale or its
flavour structure is somehow hidden behind the SM one. The second
option is more appealing. It looks more compatible with the
hierarchy problem and offers realistic chances to discover NP
directly. A natural and general proposal in this direction is the
MFV hypothesis, where the Yukawa matrices are assumed to be the only
sources of flavour violation also beyond the SM.

More precisely, we assume that there is flavour violating NP at some
high energy scale $\Lambda_{FV}$ which enters the SM lagrangian with
non-renormalizable operators suppressed by inverse powers of
$\Lambda_{FV}$. In the limit of vanishing Yukawa couplings, the
Standard Model lagrangian is invariant under the flavour symmetry
group ${\cal G}=SU(3)_{Q_{L}}\times SU(3)_{u_{R}}\times
SU(3)_{d_{R}}\times SU(3)_{L_{L}}\times SU(3)_{e_{R}}$.  The
invariance of the SM Lagrangian under ${\cal G}$ can be formally
recovered elevating the Yukawa matrices to spurion fields with
appropriate transformation properties under ${\cal G}$. The
hypothesis of MFV states that this is sufficient to make also NP
operators invariant under ${\cal G}$.

At the moment, the main motivation for extending the hypothesis of
MFV to leptons is the analogy with the quark sector, where it is
supported by a large amount of experimental data. We would like to
understand whether MFV is a general principle that acts also on
leptons. The exact analogy would imply that neutrinos are Dirac
particles. In this case, the exceptional smallness of neutrino
masses would remain unexplained and flavour changing rates like
$\mu\rightarrow e\gamma$ and $\mu - e$ conversion in nuclei would be
safely below current and future experimental limits.

For these reasons, it is more interesting to include in the MFV
hypothesis lepton number violation (LNV), that explains the
smallness of neutrino masses with the high scale of LNV,
$\Lambda_{LNV}\gg\Lambda_{FV}$. Once LNV is introduced, the concept
of analogy with quarks is ambiguous. A possible choice is to assume
that right-handed neutrinos exist (the counterpart of right-handed
up-quarks) and that their Majorana mass matrix is proportional to
the identity matrix, so that the only flavour non-diagonal objects
are the Yukawa couplings, like in the quark sector. However this
does not automatically guarantee
 that flavour changing rates are completely specified in terms of the masses and mixings
 measured
  in experiments, as it happens with quarks.

 \section{Parameter counting}\label{count}
The basic flavour changing unit we obtain from
 the spurion analysis of NP operators is
 $\lambda_\nu^\dagger\lambda_\nu$.
   In the quark sector
 the analogue quantity  $\lambda_u^\dagger\lambda_u$ can be
 expressed in terms of quark masses and CKM angles, so that the only
 unknown in the predictions for flavour changing branching ratios is the
 overall factor $1/\Lambda_{FV}^2$, coming from the power suppression of NP
 operators of dimension $>\,$4: \bea\label{qMFV} BR (quark\, MFV) \sim \frac{f(m_u,m_d, V_{CKM})}{\Lambda_{FV}^4}\eea

  Here instead we can obtain  $\lambda_\nu^\dagger\lambda_\nu$  extracting $\lambda_\nu$
 form the see-saw relation
\bea m_\nu^{\textrm eff} \equiv  v^2 \lambda_\nu^T
M_R^{-1}\lambda_\nu = \ustapmns m_{\rm diag} \udagpmns
\label{seesaw} \eea written in the basis where the charged lepton
Yukawa matrix
 $\lambda_e$ and the heavy  Majorana neutrinos mass matrix $M_R$ are diagonal.
We find \bea \lambda_\nu =
\frac{M_R^{1/2}}{v}\,O~H(\phi_1,\phi_2,\phi_3)~ m_{\rm diag}^{1/2}
~\udagpmns
 \label{lam}
\eea where $O$ is a real orthogonal matrix which depends on three
real parameters
 and $H$ a complex hermitian matrix which also depends on three real
 parameters $\phi_{1,2,3}$; $H\rightarrow I$ in the CP limit~\cite{Casas:2001sr}. In
 our definition of MFV, right-handed neutrinos are degenerate in
 mass so that $M_R$ in (\ref{lam}) is a number that we identify with $\Lambda_{LNV}$
  and the matrix $O$ can be rotated away
 (the Lagrangian
 is $O(3)$ invariant). Using eq.(\ref{lam}), the basic unit that takes part
 into low energy flavour
 changing rates in
 the MFV framework can be written as
\bea \lambda_\nu^\dagger \lambda_\nu = \frac{M_R}{v^2}\upmns\,
m_{\rm diag}^{1/2} H^2 m_{\rm diag}^{1/2}\udagpmns. \label{fc} \eea
  Let us count the parameters in
 eq.(\ref{fc}): there are 9 parameters in principle measurable at low energy (MNS angles, Dirac and Majorana phases,
 neutrino masses) and 4 unknown parameters: the normalization $M_R$ and  $\phi_{1,2,3}$ in the matrix $H$, which
 disappear in the see-saw relation (\ref{seesaw}).
With non degenerate right-handed neutrinos we would have 5 unknown
parameters more (the 2 mass splittings of $M_R$ and the 3 angles in
$O$).

Assuming that all the baryon asymmetry $\eta_B$ of the universe has
been
 generated through sphaleron effects by  a lepton asymmetry, we can use
 the observed value of $\eta_B = (6.3 \pm 0.3)\times 10^{-10}$
 to get some information on the high energy parameters $M_R$ and
 $\phi_{1,2,3}$.
  In fact, at first order
 leptogenesis depends on the combination \bea
 \lambda_\nu \lambda_\nu^\dagger = \frac{M_R}{v^2}~H~ m_{\rm diag}
  ~H \label{lept}  ~.
\eea
\section{Analysis of leptogenesis}\label{viable}
A necessary condition for generating a lepton asymmetry is the
non-degeneracy of heavy neutrinos. We in general expect that the
tree-level degeneracy of heavy neutrinos is lifted by radiative
corrections. In MFV models the most general form of the allowed
mass-splittings  is \bea \frac{\Delta M_R}{M_R}& =& c_\nu \left[
\lambda_\nu \lambda_\nu^\dagger + (\lambda_\nu
\lambda_\nu^\dagger)^T  \right] \nonumber\\ &+& c^{(1)}_{\nu\nu}
\left[  \lambda_\nu \lambda_\nu^\dagger \lambda_\nu
\lambda_\nu^\dagger +  (\lambda_\nu \lambda_\nu^\dagger \lambda_\nu
\lambda_\nu^\dagger )^T  \right] \nonumber\\&+ &c^{(2)}_{\nu\nu}
\left[ \lambda_\nu \lambda_\nu^\dagger (\lambda_\nu
\lambda_\nu^\dagger)^T \right]  + c^{(3)}_{\nu\nu}  \left[
(\lambda_\nu
\lambda_\nu^\dagger)^T \lambda_\nu \lambda_\nu^\dagger   \right]\nonumber \\
&+& c_{\nu l} \left[  \lambda_\nu \lambda_e^\dagger  \lambda_e
\lambda_\nu^\dagger +  (\lambda_\nu \lambda_e^\dagger  \lambda_e
\lambda_\nu^\dagger )^T  \right] + \ldots \label{split}\eea In a
model-independent approach we cannot fix the coefficients in
(\ref{split}) but in a perturbative scenario we expect $c_{\nu} \sim
g_{\rm eff}/4 \pi$ and $c_{\nu\nu}^{(i)},\,c_{\nu l} \sim g_{\rm
eff}^2/(4 \pi)^2$.
 From eq.(\ref{split}) we can derive some general
properties of leptogenesis in MFV models: (i) the term proportional
to $c_\nu$ does not generate an asymmetry by itself, but (ii) sets
the order of magnitude of the mass splitting and naturally gives the
condition for resonant leptogenesis: the mass splitting of
right-handed neutrinos is comparable to the decay width, \bea\Delta
M_{R_{ij}} \sim \Gamma_j = M_R
\frac{|\,\lambda_\nu\lambda_\nu^\dagger|_{jj}}{8\pi}\,.\eea  (iii)
The right amount of leptogenesis can be generated even with
$\lambda_e = 0$, if all the three parameters $\phi_{1,2,3}\neq 0$.
(iv) Since $\lambda_\nu\sim \sqrt{M_R}$, for low values of $M_R$
($\lesssim 10^{12}$ GeV) the asymmetry generated by the $c_{\nu\l}$
term dominates but is typically too small to match the observed
value of $\eta_B$. In this regime we find the flat dependence on
$M_R$ typical of resonant leptogenesis. At $M_R \gtrsim 10^{12}$ GeV
the quadratic terms $c^{(i)}_{\nu\nu}$ dominate the generation of
the asymmetry, which grows linearly with $M_R$. These specific
features of resonant leptogenesis in MFV can be derived with a
general analysis of CP-invariants and reproduced analytically.
\vspace{0.2cm}
\begin{figure}[h!]\label{fig:etaM}
\begin{center}
\includegraphics[width=4.8cm,height=5.8cm,angle=270]{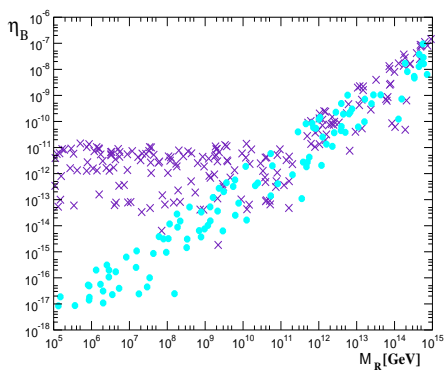}
\caption{Baryon asymmetry $\eta_B$ as a function of the right-handed
neutrinos mass $M_R$ for $c_{\nu l} = 0$ (cyan circles) and $c_{\nu
l} \neq 0$ (blue crosses).}
\end{center}
\end{figure}
 Properties
(i),(iv) explain the characteristic behaviour of $\eta_B$ as
function of $M_R$, shown in fig.~1 for reference values of the
parameters. Varying all the parameters in a wide range range we find
the result in fig.~2.
\begin{figure}[t!]\label{fig:var}
\begin{center}
\vspace{-0.5cm}
\includegraphics[width=6.6cm,height=7.3cm,angle=270]{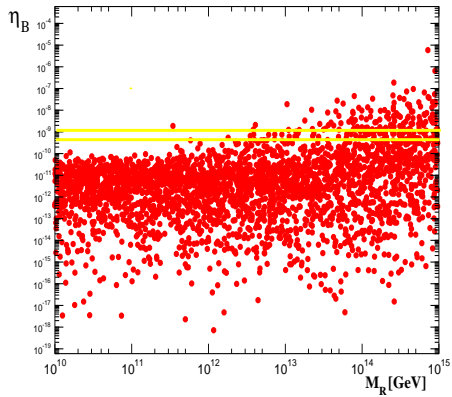}
\caption{Baryon asymmetry $\eta_B$ as a function of the right-handed
neutrinos mass $M_R$. The experimental value in log-scale is
$9.20\pm 0.02$; the larger error band takes into account the
theoretical uncertainties of a model-independent analysis, in
particular the conversion factor of the leptonic asymmetry into the
baryonic one, which depends on the field content of the model.}
\end{center}
\end{figure}
In deriving this result we used the analytic formulae for
leptogenesis without flavour effects of Ref.\cite{dibari} and
assumed a loop hierarchy between the coefficients of the mass
splittings. Under these assumptions,  the right size for $\eta_B$
can be reached for $M_R\gtrsim 10^{12}$ GeV. The regime $M_R\gg
10^{12}$ GeV is particularly interesting since in this case the
CP-violating parameters $\phi_i$ are very small and we recover the
predictive scheme of Ref.\cite{Cirigliano:2005ck}.

A MFV model  is for instance the Minimal Supersymmetric Standard
Model with degenerate right-handed neutrinos at the GUT scale. This
scenario has been recently analyzed in Ref.\cite{Branco:2006hz}. The
same behaviour shown if fig.~2 was found, but the use of a different
analysis of leptogenesis \cite{underwood} and the inclusion of
flavour effects slightly enhanced the average value of $\eta_B$ in
the low $M_R$ region so that successful leptogenesis could be
achieved for some points of the parameter space down to $10^6$ GeV.

\section{Effect of high energy parameters on $l_i\rightarrow l_j\gamma$}\label{effect}
In the MLFV framework, the effective Lagrangian relevant for the
radiative decays $l_i\rightarrow l_j\gamma$ is \bea {\cal L}_{\rm
eff}\,=\,\frac{1}{\Lambda_{\textrm LFV}^2}\,\left(c^{(1)}_{RL}
O_{RL}^{(1)}\,+\,c^{(2)}_{RL} O_{RL}^{(2)}\right)~, \eea where \bea
O_{RL}^{(1)}&\,=\,&g'H^\dagger\bar
e_R\sigma^{\mu\nu}\lambda_e\Delta_{\textrm FCNC}L_LB_{\mu\nu}~, \nonumber\\
O_{RL}^{(2)}&\,=\, &g H^\dagger\bar
e_R\sigma^{\mu\nu}\tau^a\lambda_e\Delta_{\rm FCNC}L_L W^a_{\mu\nu}~,
\eea and $g'$ ($g$) and $B_{\mu\nu}$ ($W^a_{\mu\nu}$) are the
coupling constant and the field strength tensor of the $U(1)_Y$
($SU(2)_L$) gauge group. This effective Lagrangian leads to
\bea\label{lMFV} BR_{\,l_i\rightarrow l_j\gamma} \equiv \frac{
\Gamma (\ell_i \to \ell_j \gamma) }{ \Gamma
(\ell_{i} \to \ell_{j} \nu_{i} \bar{\nu}_{j} )}\hspace{2.3cm}\nonumber\\
= 384\,\pi^2 e^2\,\frac{v^4}{\Lambda^4_{\rm
LFV}}\,\left|\left(\,\lambda_\nu^\dagger\lambda_\nu\right)_{ij}\right|^2\,
|c^{(2)}_{RL}-c^{(1)}_{RL}|^2  \nonumber  \\
\sim \left(\frac{\Lambda_{LNV} }{\Lambda^2_{\rm LFV}}\right)^2 ~ f
\left(m_{\nu}^{\rm eff}, m_l, \upmns, \phi_i \right)\hspace{0.5cm}
\label{eq:Ri2} \eea to be compared with eq.(\ref{qMFV}), which
represents a generic branching ratio in MFV in the quark sector. An
important difference is that the $\phi_i$ parameters, non-measurable
in low energy experiments, enter the branching ratio and typically
produce an enhancement. For large values of $M_R$ their effect is
moderate and MFV predicts $BR(\mu\rightarrow
e\gamma)/BR(\tau\rightarrow \mu\gamma)\lesssim 1$. A second crucial
difference between eqs.(\ref{qMFV}) and (\ref{lMFV}) is the absolute
normalization of the branching ratios, that now depends on both the
scales of lepton number and lepton flavour violation. We cannot
interpret the experimental bounds on $l_i\rightarrow l_j\gamma$ as
lower limits on the scale $\Lambda_{FV}$, without independent
information on $\Lambda_{LNV}\equiv M_R$. Leptogenesis provided this
information, since the observed value of the baryon asymmetry is
most naturally reproduced for $M_R\gtrsim 10^{12}$ GeV. For
$\Lambda_{FV}$ around the TeV, this enhances the branching ratio for
$\mu\rightarrow e\gamma$ up to values in the reach of the MEG
experiment.
\section{Summary}\label{sum}
In this talk we studied  leptogenesis in a generic MFV scenario with
right-handed Majorana neutrinos degenerate in mass. Radiative
corrections lift the tree-level degeneracy of right-handed neutrinos
and induce mass-splittings proportional to the neutrino and charged
lepton Yukawa couplings. We showed that leptogenesis is viable and
most efficient at high values of right-handed neutrino masses
($\gtrsim 10^{12}$ GeV), where it is driven by the mass-splittings
quartic in the neutrino Yukawa couplings. As a consequence,
predictions for $\mu\rightarrow e\gamma$ are enhanced and should be
observable in next experiments for natural values of the scale of
NP. High energy CP-violating  parameters, that disappear in the
see-saw relation but take part into leptogenesis, have a
significative impact on low-energy processes. The expectation
$BR(\mu\rightarrow e\gamma)/BR(\tau\rightarrow \mu\gamma)\ll 1$,
valid in the CP limit, is recovered in the regime of very heavy
right-handed neutrinos ($M_R\gg 10^{12}$ GeV).

\end{document}